\documentclass{article}

\usepackage{authblk}
\usepackage{amsmath}
\usepackage{graphicx}
\usepackage[a4paper, total={6.5in, 9.5in}]{geometry}

\begin{document}
\bibliographystyle{ieeetr}

\title{Co-existing topological and Volkov-Pankratov plasmonic edge states in magnetized graphene}
\date{}
\author[1]{Samyobrata Mukherjee \thanks{sm2575@cornell.edu}}
\author[1]{Viktoriia Savchuk}

\author[2]{Jeffery  W. Allen}
\author[2]{Monica S. Allen}

\author[1]{Gennady Shvets\thanks{gshvets@cornell.edu}}

\affil[1]{School of Applied and Engineering Physics, Cornell University, Ithaca, New York 14853, USA.}
\affil[2]{Air Force Research Laboratory, Munitions Directorate, 101 W. Eglin Blvd, Eglin AFB, Florida 32542, USA.}

\maketitle

\begin{abstract} 
Graphene placed in a perpendicular magnetic field supports optical modes known as magnetoplasmons which are transversally confined to the graphene layer. Unlike ordinary graphene plasmons, these magnetoplasmonic surface waves are characterized by a band gap corresponding to the cyclotron frequency. In addition, these magnetoplasmon bands are topological, characterized by a non-zero Chern number. This leads to the existence of topologically protected edge states at domain edges where the Chern number changes. Since the Chern number is dependent on the direction of the magnetic field, edge states exist at domain edges across which the magnetic field flips direction. Physically, the magnetic field can only flip direction at gradual domain edges with finite width creating topological heterojunctions. These topological heterojunctions support extra edge states known as Volkov-Pankratov edge states which can enter the band gap and support propagation in both directions. The number of Volkov-Pankratov states at a heterojunction varies as a function of the width of the gradual domain edge.
\end{abstract}

\section{Introduction}
Graphene is a two dimensional material comprising a monolayer of graphite with carbon atoms arranged in a honeycomb lattice. Since its isolation as a monolayer, graphene has been the subject of intensive investigation for fundamental research as well as applications \cite{Novoselov2012}. Graphene boasts attractive optical properties including a dynamically tunable optical conductivity, and strong optical nonlinearity which makes it a particularly attractive platform for photonic research and applications \cite{Graphene_Photonics_Book}. Graphene can also support surface plasmons, which are electromagnetic waves coupled to the charge carriers in graphene \cite{Graphene_Plasmonics_Book}. Graphene plasmons are deeply sub-wavelength, with low mode volumes which facilitate strong light-matter interactions, and the plasmonic spectrum of graphene can be tuned by varying the charge carrier density in the graphene layer \cite{Koppens2011, Grigorenko2012Review}. In addition, long propagation lengths have been demonstrated for graphene plasmons in high quality graphene samples \cite{Woessner2015, Alonso2017, Ni2018} and thus, graphene serves as an alternate platform for plasmonics.

While topological phases in condensed matter systems have been studied for several decades now \cite{QuantumHallReview}, the study of the topological properties of bands in optical structures is more recent. Following initial theoretical proposals \cite{Haldane2008}, topologically protected photonic states were demonstrated in tailored photonic structures including magneto-optic photonic crystals \cite{Wang2009}, photonic topological insulators using superlattices of metamaterials and pseudospin \cite{Khanikaev2013}, Floquet photonic topological insulators \cite{Rechtsman2013}, and metawaveguides \cite{Ma2015}. Topologically protected photonic states which exist at domain edges between materials with different topological invariants are particularly interesting since they are robust in the face of disorder and immune to backscattering from defects \cite{Lu2014review, Khanikaev2017review, Rechtsman2019Review}. Interestingly, edge plasmons are known to exist at the edges of graphene but they are not topologically protected \cite{Wang2011, Fei2015}. More recently, multiple approaches have been proposed to create graphene based structures that support topologically non-trivial graphene plasmons and consequently, backscattering immune graphene plasmonic edge states. While the first proposal depends on patterning the graphene to create a honeycomb superlattice \cite{Pan2017}, the second proposal involves engineering the environment of the graphene with metallic nanorods to create a SSH structure \cite{Rappoport2021}. Another scheme to realize topological graphene plasmons involves microstructuring the metallic gate adjacent to the graphene to spatially tailor the Fermi energy ($E_F$) of graphene \cite{Jung2018, Xiong2019, Fan2019, Jung2020}.

A simpler method of creating topologically non-trivial plasmonic modes would require time reversal symmetry breaking. While this may be achieved in condensed matter systems by placing them in a magnetic field \cite{QuantumHallReview, Rechtsman2019Review}, and has been demonstrated in the microwave regime for photonic states \cite{Wang2009}, it is limited in the photonic domain by the weak magneto-optical response of most materials at visible and infra-red frequencies. However, this limitation does not apply to monolayer graphene. When graphene situated in the $z=0$ plane is placed in a perpendicular magnetic field ($\vec{B}=B_0\hat{z}$), the optical conductivity becomes anisotropic and must be modeled as a tensor with non-zero values of the diagonal term (longitudinal conductivity, $\sigma_{L}$) and the off-diagonal Hall conductivity ($\sigma_{H}$) term. The magnetoplasmon band exists when $\mathrm{Im}(\sigma_{L})>0$ \cite{Ferreira2012} while $\sigma_{H}$ controls Faraday rotation \cite{Crassee2011, Ferreira2011}. The effects of shape \cite{Jiao2019, Jiao2021} and temperature \cite{Wang2012} on magnetoplasmon mode dispersion in graphene have been theoretically studied. The existence of these magnetoplasmon modes have been experimentally demonstrated via infrared spectroscopy \cite{Yan2012} as well as infrared near field optical microscopy experiments \cite{Dapolito2023}. The hybridization of graphene magnetoplasmon modes with phonon polaritons in h-BN has also been reported \cite{Wehmeier2024}.

The topological properties of magnetoplasmon bands in 2D electron gas (2DEG) systems has been studied \cite{Jin2016}. It was found that the magnetoplasmon band supported by a 2DEG system placed in a perpendicular magnetic field has non-trivial topology characterized by a non-zero Chern number, $C=\pm 1$, whose sign depends on the direction of the magnetic field. Domain edges where the magnetic field flips sign was found to support two topologically protected edge states while domain edges  where the 2DEG terminates was found to support one topologically protected edge state. The existence of kink magnetoplasmons at domain edges where the magnetic field flips direction has been experimentally verified in a GaAs/AlGaAs 2DEG \cite{Jin2019}. For graphene in particular, the first experimental evidence of backscattering being suppressed in graphene magnetoplasmonic edge states was reported in Ref. \cite{Yan2012}. The existence of topologically protected edge states at domain edges has been predicted for periodically patterned graphene placed in a magnetic field \cite{Jin2017} as well as for a graphene monolayer screened by a gate \cite{Ciobanu2022}.

In this article, we consider free-standing graphene placed in a perpendicular magnetic field which can support magnetoplasmon modes. Applying a magnetic field results in a band gap and we find that the magnetoplasmon bands supported by graphene have non-trivial topology and non-zero Chern numbers can be assigned to these bands by calculating the Berry phase. We focus our study on edge states at domain boundaries where the magnetic field flips direction. From the bulk-interface correspondence, two topological edge states which span the band gap are expected since the magnetoplasmon band on either side of the domain edge has opposite signs of the Chern number $C=\pm 1$ \cite{Tauber2019}. However, sharp transitions in the direction of the magnetic field from $+B_0\hat{z}$ to $-B_0\hat{z}$ are unrealistic since physical systems where the magnetic field flips direction involve length scales ranging from microns for setups involving microstructured ferromagnetic materials \cite{Ye1995, Nogaret2000} to $\approx 0.5$ mm for a NdFeB magnet \cite{Jin2019}. The domain edge where the magnetic field flips direction over a finite length $L$ thus constitutes a topological heterojunction and we find that increasing $L$ leads to the appearance of new massive edge states in addition to the two topological edge states predicted by the bulk-interface correspondence. These new edge states are known as Volkov Pankratov (VP) states, named after the physicists who first predicted their existence \cite{Volkov1985} and are a general feature of topological heterojunctions where they appear for domain boundaries of finite, non-zero width. VP edge states have been predicted at topological heterojunctions involving topological insulators \cite{Tchoumakov2017}, topological superconductors \cite{Alspaugh2020}, topological graphene ribbons with intrinsic spin orbit coupling \cite{Van2020} and have been experimentally demonstrated at topological heterojunctions involving mercury telluride \cite{Mahler2019} and lead tin selenide \cite{Bermejo2023}. While VP edge states have been reported in electronic systems, to the best of our knowledge, this is the first report of the existence of VP edge states at topological heterojunctions in photonic (specifically, graphene plasmonic) systems.

\section{Theoretical Model}
\subsection{Magneto-optical conductivity of graphene}
\begin{figure}[t]
\centering\includegraphics[width=1.0\linewidth]{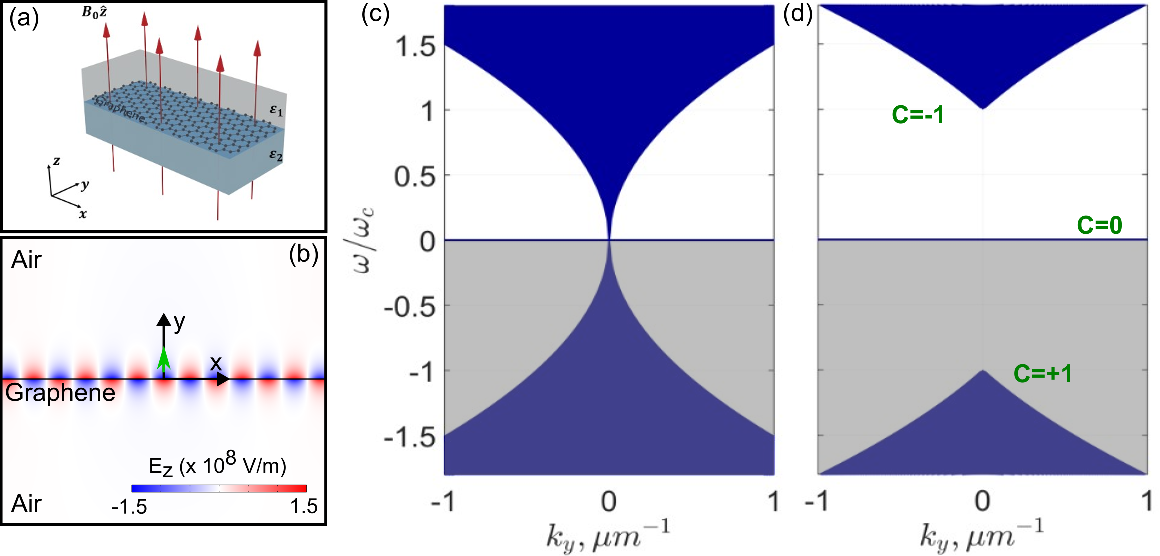}
\caption{(a) Schematic showing a graphene monolayer lying in the $z=0$ plane placed in an external magnetic field ($\vec{B}=B_0\hat{z}$). (b) Normal electric field component $E_z$ for a bulk graphene magnetoplasmon excited by a point dipole source (green arrow) at frequency $\omega=1.408\omega_c$. (c) Bulk spectrum for graphene plasmons $(E_F = 70\ \rm{meV)}$ as the function of $k_y$ when $B_0=0$. (c) Same as (b) but for $B_0  = 1\ T$. Chern numbers for the bulk bands are denoted in green.}
\label{fig:1}
\end{figure}

We study a free-standing graphene layer situated in the $z=0$ plane between two  dielectrics (with permittivity $\epsilon_1$ and $\epsilon_2$) filling the semi-infinite half-spaces $z>0$ and $z<0$, respectively, with a uniform magnetic field ($\vec{B}=B_0\hat{z}$) applied perpendicular to the graphene as shown in Fig. \ref{fig:1} (a). The electronic band structure of graphene placed in a perpendicular magnetic field condenses into highly degenerate Landau levels with energies $E_n=sign(n)v_F\sqrt{2\hbar e B_0|n|}$, where $v_F$ is the Fermi velocity in graphene, and $B_0$ is the magnitude of the magnetic field. Generally, modeling the full magneto-optical conductivity for graphene requires the consideration of both intra-band and inter-band electronic transitions between these Landau levels.

However, in this article we are interested in the low frequencies where the effects of the intra-band transition dominate and we neglect the higher energy inter-band transitions. We also restrict ourselves to the local magneto-optical conductivity of graphene where we neglect its dependence  on the in-plane wavevector $\vec{k}$. Under these assumptions, the magneto-optical conductivity of graphene may accurately be modeled as a semi-classical Drude response \cite{Ferreira2011, Ferreira2012} with the additional requirement that multiple Landau levels are fully filled. We find that having at least three fully filled Landau levels ($N_F>3$) yields reasonable agreement between the semi-classical Drude conductivity and the full magneto-optical conductivity at low frequencies. The requirement of $N_F>3$ requires that the Fermi energy $|E_F|>v_F \sqrt{6\hbar e B_0}$. The low frequency regime is defined as $\omega\ll \omega_{inter}$, where $\omega_{inter}=v_F\sqrt{\frac{2eB_0}{\hbar}}\{2+\sqrt{3}\}$ corresponds to the first allowed inter-band transition (between the third Landau level in the valence band and the fourth Landau level in the conduction band, for negatively doped graphene) for the case with $N_F>3$. For example, for $B_0=1\ \rm{T}$ and $v_F=10^6 \ \rm{m/s}$, the Drude conductivity is valid for $E_F>62.86\ \rm{meV}$ and $\omega\ll 205.72\times 10^{12}\ \rm{rad/s}$ $(\rm{f}\ll 32.74\ \rm{THz})$. The Drude conductivity contains longitudinal ($\sigma_L$) and Hall ($\sigma_H$) terms given as
\begin{equation}
    \begin{split}
    \sigma_L(\omega)&=\frac{ie^2E_F}{\pi\hbar^2}\cdot\frac{\omega+i\gamma}{(\omega+i\gamma)^2-\omega_c^2}, \\
    \sigma_H(\omega)&=\frac{e^2E_F}{\pi\hbar^2}\cdot\frac{\omega_c}{(\omega+i\gamma)^2-\omega_c^2},
    \end{split}
    \label{eq:sigma}
\end{equation}
where $\omega_c=\frac{eB_0v_F^2}{E_F}$ is the intra-band cyclotron frequency and $\gamma$ is the scattering rate.

\subsection{Governing equations and band structure}

The dynamics of surface magnetoplasmons in graphene can be captured by the continuity equation and a constitutive equation including the potential defined in the $z=0$ plane \cite{Ferreira2011, Jin2016}:
\begin{equation}
    \label{governing_equations}
    \begin{split}
        \omega\rho(\vec{r},\omega) &= -i\nabla \cdot \vec{j}(\vec{r},\omega), \\
        \omega\vec{j}(\vec{r},\omega) &= -i \frac{e^2E_F(\vec{r})}{\pi\hbar^2}\nabla\phi(\vec{r},\omega) - i\omega_c(\vec{r})\vec{j}(\vec{r},\omega)\times \hat{e}_z,
    \end{split}
\end{equation}
where $\rho(\vec{r},\omega)$ is a small perturbation of the charge density off its equilibrium value, $\vec{j}(\vec{r},\omega)$ is the 2D surface current density in the $z=0$ plane, $E_F(\vec{r})$ is the Fermi energy, and $\omega_c(\vec{r})$ is the space dependent cyclotron frequency.
The in-plane scalar potential in momentum space then takes the form
\begin{equation}
\label{eq:potential}
\phi(\vec{k})=\frac{\rho}{(\epsilon_1+\epsilon_2)\epsilon_0k},
\end{equation}
where $k=\sqrt{k_x^2+k_y^2}$.

\subsubsection{Bulk states}
When considering the bulk spectrum where the magnetic field is invariant in the $z=0$ plane, a Fourier transform of the in-plane spatial coordinates leads to $\nabla \rightarrow i\vec{k}$ and the governing equations (Eq. \ref{governing_equations}) may be rewritten as a Hermitian eigenvalue problem, $H|\psi\rangle=\omega|\psi\rangle$, where $|\psi\rangle = (\tilde{ \rho}\equiv\frac{\alpha}{\sqrt{k}} \rho, j_x,j_y)^T$ and 
\begin{equation}
\label{hamiltonian}
H = 
\begin{pmatrix}
0&\alpha \frac{k_x}{\sqrt{k}}&\alpha \frac{k_y}{\sqrt{k}}\\
\alpha \frac{k_x}{\sqrt{k}}&0&-i\omega_c\\
\alpha \frac{k_y}{\sqrt{k}}&i\omega_c&0
\end{pmatrix},
\end{equation}
where $\alpha = \sqrt{\frac{e^2E_F}{2\pi\hbar^2\epsilon_0}}$ and we assume that $\epsilon_1=\epsilon_2=1$ for free standing graphene. Solving this eigenvalue problem yields the spectrum
\begin{equation}
\label{eq:bulk_spectrum}
    \begin{split}
        \omega_\pm(\vec{k})&=\pm \sqrt{\omega_c^2+\alpha^2 k}, \\
        \omega_0(\vec{k})&=0, 
    \end{split}
\end{equation}
where the subscripts $+/-/0$ represent the positive, negative, and zero frequency bands, respectively. The corresponding non-normalized eigenvectors are
\begin{equation}
\label{eigenfunction_plus_minus}
|\psi_{\pm}\rangle=
\begin{pmatrix}
\alpha^2k\\
\pm\alpha\frac{k_x}{\sqrt{k}}\sqrt{\omega_c^2+\alpha^2k}-i\omega_c\alpha\frac{k_y}{\sqrt{k}}\\
\pm\alpha\frac{k_y}{\sqrt{k}}\sqrt{\omega_c^2+\alpha^2k}+i\omega_c\alpha\frac{k_x}{\sqrt{k}}
\end{pmatrix},\  
|\psi_{0}\rangle=
\begin{pmatrix}
\omega_c\\
i\alpha\frac{k_y}{\sqrt{k}}\\
-i\alpha\frac{k_x}{\sqrt{k}}.
\end{pmatrix}.
\end{equation}

\subsubsection{Chern Numbers}

In the case of continuous media, one can map the $(k_x, k_y)$ plane into the Riemann sphere \cite{Silveirinha2015,Parker2021}. The Chern number is defined as the flux of the Berry curvature

\begin{equation}
\label{chern_number}
C_{\pm}=\int (\nabla_k\times\vec{A}_\pm)d^2\vec{k}
\end{equation}
where $\vec{A}_\pm=-\frac{\mathrm{Im}\langle\psi_\pm|\nabla_k\psi_\pm\rangle}{\langle\psi_\pm|\psi_\pm\rangle}$ is the Berry connection. Following some algebra, we can define the Chern number for the bulk magnetoplasmon bands in graphene as 
\begin{equation}
\label{eq:chern_number}
C_{\pm}=\mp \int\frac{\alpha^2\omega_c}{2(\omega_c^2+\alpha^2k)^\frac{3}{2}}d\vec{k}=\mp \mathrm{sign}(\omega_c),\ 
C_{0}=0.
\end{equation}

\subsubsection{Edge States}
Whereas the magnetoplasmon dispersion for a graphene layer placed in a perpendicular magnetic field can be obtained from Eq. \ref{eq:bulk_spectrum}, this solution cannot account for any spatial variation in the magnetic field. We use the plane wave expansion (PWE) method to address the situation with a spatially varying magnetic field, specifically, with the magnetic field varying along $x$ but invariant along $y$ \cite{Jin2016}.  We consider three magnetic field domains in our calculation window with $\vec{B}= B_0\hat{z}$ in the central (purple) domain and $\vec{B}=-B_0\hat{z}$ in the outer (brown) domains as shown in Fig. \ref{fig:2}(a). Thus, we have magnetic domains as strips running parallel to $\hat{y}$. The width of the central domain is $W$ and the width of the computational window is $P=2W$. The spatial variation of the magnetic field is encapsulated in a spatially varying $\omega_c(x)$ in Eq. \ref{governing_equations} and we expand it in the basis of plane waves along $x$ as
\begin{equation}
    \omega_c(x)=\sum_{n=-N}^N\tilde{\omega}_{c,n}e^{ik_{x,n}x}
\end{equation}
where $N$ is the number of plane waves used in the expansion and $k_{x,n}=\frac{2\pi}{P}\cdot n$, with $P$ being the size of the computation window. Similarly, we also expand $j_x(x), j_y(x)$ and $\rho$. Since we work with a finite number of waves in the PWE method, we expand our terms in a Fourier series and this implies a periodicity $P$. Therefore, the dispersion plots calculated with the PWE method are for a periodic array of domains of width $W$ with opposite magnetic field, with period $P$. However, the periodicity does not affect the dispersion of the edge states since they are localized around the domain edges and generally do not extend to the ends of the computation window.

Following some algebra and using the equation for $\phi(k)$ provided in Eq. \ref{eq:potential}, we formulate an eigenvalue equation which yields the dispersion plot $\omega(k_y)$ for the bulk modes and edge states. The eigenvectors are obtained as functions of $k_x$ and have to be inverse Fourier transformed to obtain $\rho, j_x$ and $j_y$ as functions of $x$. Ideally, $N$ should be $\infty$ but practically we choose N=200, leading to a total of 401 waves being used in the calculation. While the choice of a finite $N$ limits the accuracy of this method in modeling sharp variations in the magnetic field profile, this method can be used to model smooth changes in the magnetic field profile accurately. Another possibility to model gradual edge profiles would be to solve for the potential using an integral equation \cite{Wang2011}.

\section{Results and Discussion}

\subsection{Bulk states}
We assume that the graphene is lossless ($\gamma\to0$) and consider $B_0=1$ T, $E_F=70 \ \rm{meV}$ and $v_F=10^6\ \rm{m/s}$ for our numerical calculations unless otherwise specified. For these parameters $f_c=\frac{\omega_c}{2\pi}\approx2.27$ THz. We calculate the eigenvalues of the Hamiltonian given in Eq. \ref{hamiltonian} to obtain the bulk magnetoplasmon band spectrum. Solving with $B_0=0$ yields the classic graphene plasmon dispersion with $\omega\propto\sqrt{k}$. However, in the presence of a uniform magnetic field, $B_0=1$ T, we find 3 bulk magnetoplasmon bands with a band gap equaling $\omega_c$ between the bulk bands at $\omega=0$ and at positive and negative frequencies. Moreover, calculating the Berry phase, we find that the bulk band at positive (negative) frequencies has a non-zero Chern number $C=-(+)1$. These properties are similar to magnetoplasmons of two-dimensional electron gas systems (2DEG) reported in Ref. \cite{Jin2016}. However the shape of the graphene magnetoplasmon band is different from that of the 2DEG magnetoplasmon band, specifically at $k_y=0$ where the graphene magnetoplasmon exhibits a sharp transition unlike the parabolic shape seen for magnetoplasmon bands in 2DEGs. Moving forward, we shall ignore the bands at negative frequencies in the remainder of this article.

\begin{figure}[t]
\centering\includegraphics[width=1.0 \linewidth]{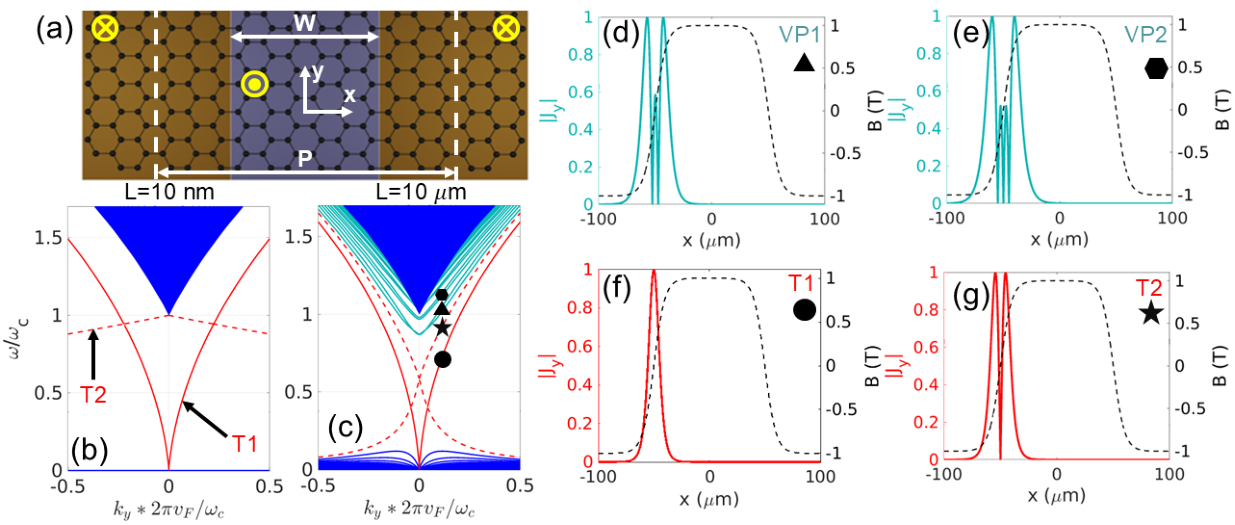}
\caption{(a) Schematic showing the domain edges where magnetic field flips direction. (b) Dispersion plot for relatively sharp domain edges where $L_1=L_2=10$ nm. The bulk band is shown in blue while the two topological edge states are shown in red (T1 - solid, T2 - dashed) for $B_0=1$ T, $E_F=70\ \rm{meV}$. (c) Same as (b) but for much smoother domain edges with $L_1=L_2=10$ $\mu m$. (d)-(g) The edge state profiles for the left domain edge showing $|J_y|$ for the two topological edge states and the first two VP states at $k_y=0.1 \cdot \omega_c/ (2\pi v_F)$. The dashed black line shows the magnetic field profile. The black shapes serve as guides for the eye.}
\label{fig:2}
\end{figure}

\subsection{Topological and Volkov Pankratov edge states  }
We now focus on a system with the magnetic field of equal magnitude ($B_0=1\ \rm{T}$) pointing in opposite directions in different parts of our calculation domain as shown in Fig. \ref{fig:2}(a). We choose the width of the computation window $P=200\ \rm{\mu m}$, and the width of the magnetic domain strip $W=100\ \rm{\mu m}$. In order to investigate the edge states that exist at domain edges of finite width, we model the magnetic field profile as
\begin{equation}
\label{eq:B_profile}
\vec{B}(x)=B_0\cdot\tanh\left({\frac{x-x_1}{L_1}}\right)\cdot \tanh\left(-{\frac{\pm x-x_2}{L_2}}\right)\hat{z},
\end{equation}
where $x_{1}, x_2$ are the positions of the domain edges and $L_1$, and $L_2$ give the length scale over which the magnetic field flips direction at the respective domain edges. We choose $x_{1}=-50\ \rm{\mu m}$ for the rising magnetic field ($-B_0\to+B_0$) domain edge and $x_0=+50\ \rm{\mu m}$ for the falling ($+B_0\to-B_0$) edge.

Since the Chern number of the bulk magnetoplasmon band is $C=\pm1$ depending on the direction of the magnetic field, the difference in the Chern number across the domain edge is $2$. Thus from the bulk-interface correspondence, we expect two topological edge states that span the band gap. This is what we find when we use the PWE method to obtain the dispersion plot for a relatively sharp domain edge with $L_1=L_2=10 \rm{nm}$ as shown in Fig. \ref{fig:2}(b). There are fast (high group velocity $v_g$) and slow (low group velocity $v_g$) topological edge states that span the band gap at each domain edge. The fast (T1) and slow (T2) topological edge states are shown in Fig. \ref{fig:2}(b) using solid and dashed red lines, respectively. The rising, $x_1=-50\ \rm{\mu m}$, (falling, $x_2=50\ \rm{\mu m}$) domain edge houses the forward (backward) propagating edge states. The forward (backward) propagating slow edge state has negative (positive) phase velocity and only exists for the domain where $k_y<0\ (k_y>0)$. Each topological edge state that is confined to a domain edge only propagates in one direction. This is the property that ensures unidirectional propagation of topological edge states making them immune to backscattering.

There are several changes to the dispersion of the graphene magnetoplasmon bulk and edge states when we move to a smoother domain edge with $L_1=L_2=10\ \rm{\mu m}$ as shown in Fig. \ref{fig:2}(c). Firstly, we find that the $|v_g|$ of the "slow" topological edge states has increased and become comparable to the $|v_g|$ of the "fast" topological edge states. We shall continue to refer to the "fast" topological edge state which has a lower frequency and passes through $\omega=0$ at $k_y=0$ as T1 (solid red line in Fig. \ref{fig:2}(c)) and refer to the "slow" topological edge state which exists at higher frequencies and passes through the middle of the band gap ($\approx 0.57\omega_c$) as T2 (dashed red line in Fig. \ref{fig:2}(c)). At $k_y=0$, T2 merges with the bulk band at $\omega=\omega_c$ for the sharp domain edge but moves towards the center of the bandgap, appearing at $\omega\approx0.57\cdot\omega_c$ for this smooth domain edge. Moreover, the forward (backward) propagating T2 edge state now exists in the domain where $k_y>0\ (k_y<0)$. In addition, we find that several new edge states, shown in cyan in Fig. \ref{fig:2}(c) have appeared outside the bulk bands. These are Volkov Pankratov (VP) edge states that appear as a consequence of the finite width of the topological heterojunction created by the gradual flipping of the direction of the magnetic field at a smooth domain edge \cite{Tchoumakov2017}. Although the VP states enter the band gap, they do not span the gap. The VP states appear in pairs and their propagation is not unidirectional. We find that each VP edge state has positive (negative) $v_g$ when $k_y>0\ (k_y<0)$. Thus, a VP edge state that encounters a scatterer along its propagation path can be scattered into a backward propagating mode at the same frequency but a different value of $k_y$. Thus, VP states, while a consequence of a topological heterojunction, differ from topologically protected edge states in that they are not immune to backscattering.

The transverse profiles of some of the edge states, specifically, the tangential component of the surface current density $J_y$ is plotted in Fig. \ref{fig:2}{d-g} for $k_y=0.1 \cdot \omega_c/ (2\pi v_F)$. We find that the two topological edge states resemble the fundamental and first order modes of a waveguide, while moving to the VP edge states at higher frequencies reveals transverse profiles with more nodes. The two lowest frequency VP edge states (VP1 and VP2) at the $x=-50\ \rm{\mu m}$ domain edge, with two and three nodes, respectively, are shown in Fig. \ref{fig:2}(d) and (e). The VP edge states are also wider than the topological edge states. It is useful to consider the gradual domain edge as a waveguide. The two topological edge states T1 (no nodes) and T2  (one node) may be considered as the fundamental and first-order mode, respectively, with more nodes (half-wavelengths) being added for higher order VP edge states \cite{Wang2011}. The magnitude of the magnetic field gradually decreases to zero (before increasing again)) at the domain edge. For zero magnetic field, there is no cyclotron frequency and therefore, no cutoff for this interfacial waveguide. Therefore, the VP states which exist only at the domain edge where the magnetic field has low magnitude, can exist at frequencies below $\omega_c$. Note that the edge state dispersion is not particularly sensitive to the exact profile of the edge state  as defined in eq. \ref{eq:B_profile} and similar results may be obtained using linear profiles. The edge state dispersion, however, is sensitive to the width of the edges $L$.

\subsection{Excitation of edge states}
\begin{figure}[t]
\centering\includegraphics[width=1.0 \linewidth]{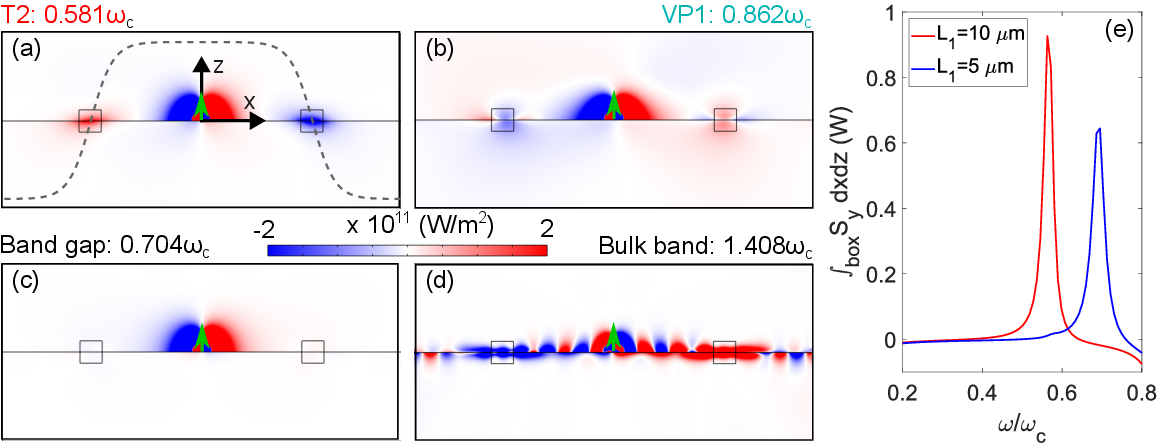}
\caption{(a) Driven simulation showing Poynting flux into the plane $S_y$ for a point dipole source oriented along $+\hat{z}$ placed at $(0,-4)\ \rm{\mu m}$ (green arrow) with angular frequency $\omega=0.581\omega_c$. The graphene is placed at $z=0$ (solid black line). The dashed gray line shows the magnetic field profile ($B_0=1$ T, $x_1=-50\ \rm{\mu m},L_1=10\ \rm{\mu m}$ and $x_2=50\ \rm{\mu m}, L_2=9.5\ \rm{\mu m}$). The small boxes serve as guides to the eye showing where the magnetic field domain edges lie. (b), (c), and (d) are the same as (a) but for $\omega=0.862\omega_c$, $\omega=0.704\omega_c$, and $\omega=1.408\omega_c$ respectively. (e) Integral of $S_y$ in the small box but only for the section below the graphene ($z<0$) for the rising edge at $x_1=-50\ \rm{\mu m}$ with $L_1=10\ \rm{\mu m}$ (red) and $L_1=5\ \rm{\mu m}$ (blue) as a function of $\omega$.}
\label{fig:2_5}
\end{figure}

We use a commercially available finite element solver, COMSOL Multiphysics, to study the excitation of the magnetoplasmonic edge states. We set a computational domain of $200 \ \rm{\mu m}$ in the $x$ direction and $100\ \rm{\mu m}$ in the $z$ direction with perfectly matched layers (PML) of $10\ \rm{\mu m}$ thickness at the edges of the computation domain. The graphene layer is placed at $z=0$ and is characterized by its optical conductivity given in Eq. \ref{eq:sigma}. However, unlike the dispersion calculations where graphene was assumed to be lossless, we assume a finite scattering rate $\gamma=0.1$ THz for the driven simulations. This leads to the ratio $\gamma/\omega_c=0.007$ which results in a slight blurring of the band edges while still maintaining a band gap and the edge states \cite{Jin2019}. Higher values of $\gamma$ may lead to lesser localization of the edge states and in extreme cases may even lead to the band gap effectively closing due to bands being smoothed out and the density of states in the band gap becoming non-zero. A point dipole source is placed at the center of the computational domain ($x=0$) at a distance of $4\ \rm{\mu m}$ from the graphene layer ($z=4\ \rm{\mu m}$) as shown in Fig. \ref{fig:2_5}(a) which shows the computation domain sans the PML. The magnetic field profile is similar to the one used for the calculation of the dispersion plot in Fig. \ref{fig:2}(c) but we break symmetry about $x=0$ to aid visualization. The magnetic field profile is given by Eq. \ref{eq:B_profile} with $B_0=1$ T, $x_1=-50\ \rm{\mu m},L_1=10\ \rm{\mu m}$ and $x_2=50\ \rm{\mu m}, L_2=9.5\ \rm{\mu m}$ and is shown schematically in Fig. \ref{fig:2_5}(a). The structure is assumed to be invariant along the $y$ direction and we choose $k_y=0$.

Fig. \ref{fig:2_5}(a) shows the result of the driven simulation at $\omega=0.581\omega_c$, the frequency corresponding to the "slow" topological edge state T2 at $k_y=0$. The Poynting flux in the direction perpendicular to the plane ($S_y$) is shown. The dipole at $(0,4)\ \rm{\mu m}$ excites an edge state propagating along $+\hat{y}$ ($-\hat{y}$) at the rising (falling) edge at $x=-50\ \rm{\mu m}$ ($x=+50 \mu m$). The edge states decay away from the magnetic field domain edges and also away from the graphene. Similarly, a dipole source at $\omega=0.862\omega_c$ excites the VP1 state that exists at that frequency in Fig. \ref{fig:2_5}(b). $S_y$ for the VP1 state at each edge, while lower in magnitude is opposite in sign compared to T2. This is consistent with the lower group velocity at $k_y=0$ for the slightly asymmetric magnetic field profile. These edge states are excited by evanescent fields from the dipole source. Gradual domain edges support edge states that are less tightly localized compared to the edge states at sharp domain edges and this makes it easier to excite edge states at gradual domain edges via evanescent coupling. Choosing a frequency in the band gap away from the frequency at which the edge states exist results in no edge state being excited at the domain edges as shown in Fig. \ref{fig:2_5}(c) for $\omega=0.704\omega_c$. Choosing a frequency within the bulk bands ($\omega=1.408\omega_c$) leads to the excitation of a bulk magnetoplasmon that decays away from the graphene in the $z$ direction in all the magnetic field domains along the graphene layer placed at $z=0$.

We further study the impact of variation of the domain edge thickness on the excitation of the edge states, focusing the on the topological edge state T2. We focus on the flux $S_y$ within the small box at the rising edge (centered at $x=-50\ \rm{\mu m}, z=0$) shown in Figs. \ref{fig:2_5}(a)-(d). We only consider the flux in the area below the graphene layer ($z<0$) to avoid any distortions in the flux due to the dipole source placed in the $z>0$ domain which are mostly screened by the graphene layer. Fig. \ref{fig:2_5}(e) shows $\iint S_y dx dz$ as a function of the excitation frequency for two different values of $L_1$. For $L_1=10\ \rm{\mu m}$, we see that there is a peak in the flux at the edge centered at the frequency of the T2 edge state. A sharper domain edge with $L_1=5\ \rm{\mu m}$ results in the flux peak shifting to a higher frequency, as expected from the fact that for a sharp edge, T2 exists at $\omega_c$. The impact of the domain edge thickness on the frequency of edge states and the existence of VP states is studied in more detail in the next section.

\subsection{VP edge states as a function of edge thickness}

\begin{figure}[t]
\centering\includegraphics[width=0.4 \linewidth]{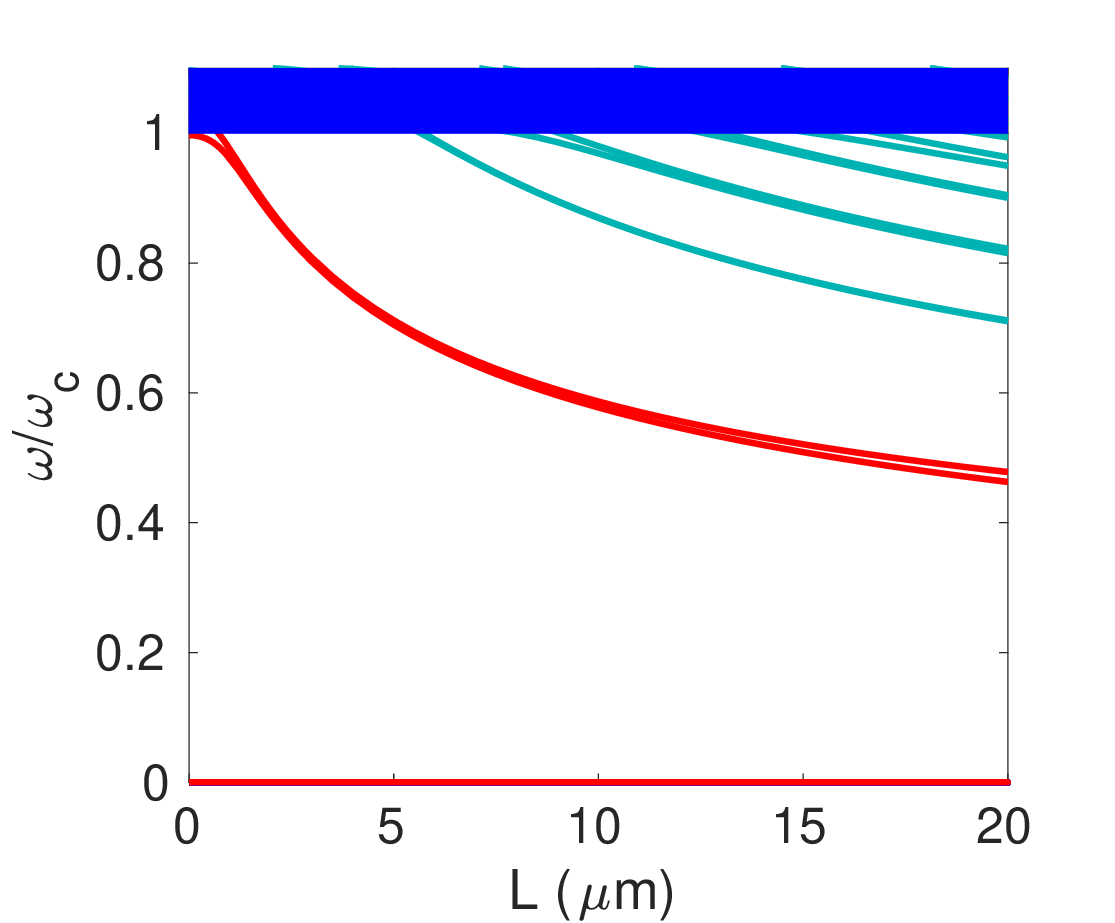}
\caption{Edge state dispersion as a function of the width of the topological heterojunction, $L_1=L_2=L$ at $k_y\to 0$. Topological edge states are shown in red while the VP edge states are shown in cyan and bulk modes are shown in blue.}
\label{fig:3}
\end{figure}

 We plot the VP states at $k_y\to0$ as a function of the domain edge width $L$ in Fig. \ref{fig:3}. We find that with $k_y\to0$, the "fast" topological edge state T1 (bottom red line in Fig. \ref{fig:4}) exists at $\omega=0$ regardless of the value of L. T1 coexists with the $\omega=0$ bulk band. The "slow" topological edge state T2 exists at $\omega=\omega_c$ when $L\to0$ but moves to lower frequencies as $L$ increases. The number of VP edge states varies as a function of the width of the topological heterojunction at the domain edge. While there are no VP edge states at $k_t=0$ for low values of $L$, VP edge states appear as $L$ increases. The VP edge states originate in the $\omega>\omega_c$ bulk band and move to lower frequencies as $L$ increases. The VP edge states appear in pairs corresponding to the two domain edges in our computation window. For very smooth edges at large values of $L$, the edge states are not strongly localized at the domain edges and this leads to edge states from the two domain edges hybridizing with each other. This coupling  explains the gradual separation of the T2 edge states for $L>15\ \rm{\mu m}$.

\subsection{Unequal magnetic field and edge state coupling}
\begin{figure}[t]
\centering\includegraphics[width=1.0 \linewidth]{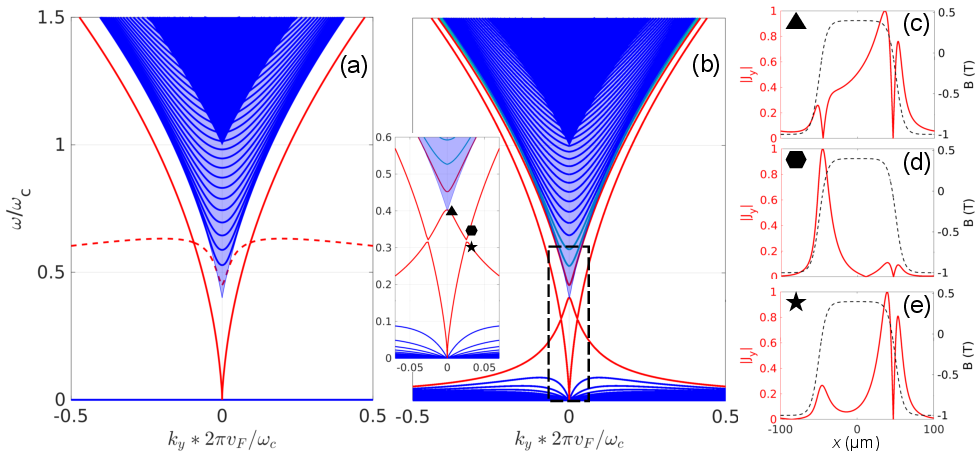}
\caption{(a) Dispersion plot for $L=10$ nm for unequal magnetic fields pointing in opposite directions. The magnetic field in the central domain is $0.4$ T along $+\hat{z}$ while the field in the outer domains is  $1$ T along $-\hat{z}$. The translucent blue band shows the part of the band that corresponds to the bulk for $B_0=0.4$ T , while the solid blue band denotes bulk modes for both domains ($B_0=0.4$ T as well as $B_0=1$ T). The red lines show the topological edge states T1 (solid) and T2 (dashed). (b) Same as (a) but for smooth domain edges with $L=10\ \rm{\mu m}$. The insets shows a zoomed in picture of the band gap and edge states around $k_y=0$. (c)-(e) shows the edge state profiles ($|J_y|$) in red at the points on the edge state dispersion branches marked with the black shapes. The dashed black line shows the magnetic field profile.}
\label{fig:4}
\end{figure}

We find that unequal magnitudes of the magnetic field pointing in opposite directions across a domain edge reveals new properties of the edge states including the possibility of tuning the coupling or hybridization of the edge states. We now consider the a field of $0.4$ T along $+\hat{z}$ in the central domain ($-50\ \rm{\mu m}\leq x \leq +50\ \rm{\mu m}, W=100\ \rm{\mu m}$) while the field in the outer domains is  $1$ T along $-\hat{z}$. Such a configuration could be experimentally realized using multi-magnet setups including permanent micro-magnets and a tunable external magnetic field provided by an electromagnet \cite{Jin2019}.  Since the Chern number is only a function of the orientation of the magnetic field and not its magnitude, we again expect two topological edge states at each interface from the bulk-interface correspondence. However, two different magnitudes of the magnetic field in the domains result in two different values of $\omega_c$ for the different magnetic field domains. This is shown in Fig. \ref{fig:4} (a) and (b) with the translucent blue area showing the bulk band that corresponds only to the domain with $0.4$ T. More interestingly, we find that in the case with relatively sharp domain edge, $L=10$ nm, the "slow" topological edge state T2 (dashed red line in Fig. \ref{fig:4}(a)), on the left and right domain edges are connected via the bulk band for the lower magnetic field domain. This is different from the case with equal magnetic field pointing in opposite directions, where T2 merges with the bulk band at $\omega=\omega_c$. Increasing $L$ to $10\ \rm{\mu m}$ results in a drastic transformation of the graphene magnetoplasmon dispersion. While some VP edge states appear at higher frequencies, we notice band gaps opening due to hybridization between edge states from opposite domain edges. The lower cyclotron frequency of the central domain with $\vec{B}=0.4$ T $\hat{z}$ favors coupling between edge states from opposite domain edges through that domain. We find that near $k_y=0$, the two "slow" topological edge states T2, which exist on opposite domain edges, have coupled through the central domain as shown in Fig. \ref{fig:4}(c) and that hybridization has led to the appearance of a band gap. A smaller band gap also appears at $k_y\approx0.026 \cdot \omega_c/ (2\pi v_F)$, due to coupling between the "fast" (T1) and "slow" (T2) topological edge states on opposite domain edges. The edge state profiles above and below this band gap at $k_y=0.0265\cdot \omega_c/ (2\pi v_F)$ are shown in in Fig. \ref{fig:4} (d) and (e) which correspond to the symmetric and anti-symmetric coupled mode profiles.

\section{Conclusion}
We have studied the topological properties of the magnetoplasmon bands in graphene. We have found that free-standing graphene placed in a perpendicular magnetic field can support topologically non-trivial surface magnetoplasmons with Chern numbers of $\pm1$ depending on the direction of the magnetic field. Upon investigating the edge states at the interface between two magnetic domains, we found that in addition to the topological edge states predicted by the bulk-interface correspondence, new edge states appear when the domain edge is smooth instead of sharp, creating a topological heterojunction. The new edge states are graphene magnetoplasmonic Volkov Pankratov edge states that have been known to appear at topological heterojunctions in electronic systems. Larger numbers of VP edge states appear in the band gap as the domain edge (topological heterojunction) is made wider. We also found that unequal magnitudes of the magnetic field in different domains lead to favorable circumstances for the hybridization of edge states from opposite domain edges via the domain with lower magnetic field. The control over coupling via the magnetic field could lead to potential applications of these edge states in building tunable THz interferometers using graphene based platforms.

\section*{Acknowledgements}
The work at Cornell was supported by the University of Dayton Research Institute (UDRI) under the contract FA8651-24-F-B013, Office of Naval Research (ONR) under the grant no. N00014-21-1-2056, and the Army Research Office (ARO) under the award W911NF2110180. Authors JWA and MSA would like to acknowledge support from the CLAWS Applied Research for the Advancement of Priorities program of the Office of the Secretary of Defense.

\section*{Disclosures}
The authors declare no conflicts of interest.

\section*{Data Availability}
No experimental data was generated for the research presented in this article.

\bibliography{Magnetoplasmons}

\end{document}